\pdfoutput=1  

\documentclass[aps,prd,superscriptaddress,floatfix,nofootinbib,preprintnumbers,eqsecnum,twocolumn]{revtex4-1}

\usepackage[normalem]{ulem}
\usepackage{amsmath,amssymb,float,graphicx,placeins,multirow}
\usepackage{enumerate,slashed,cancel,comment,color,bbm,rotating,placeins,mathtools,multirow,hhline}
\usepackage{array}
\usepackage{empheq}
\usepackage{xkcdcols} 
\usepackage{hyperref}
\usepackage{lipsum}
\hypersetup
{
  colorlinks,
  linkcolor={blue!80!black},
  citecolor={blue!80!black},
  urlcolor={blue!80!black}
}

\usepackage{mymacros}

\usepackage{soul} 

\graphicspath{{./figs}}

\begin{document}

\title{
Machine Learning Does It
and Does It Better:\\
Unearthing Primordial Dark-Matter Velocities
from the Matter Power Spectrum}

\def\andname{\hspace*{-0.5em}} 
\author{Keith R. Dienes}
\email[Email address: ]{dienes@arizona.edu}
\affiliation{Department of Physics, University of Arizona, Tucson, AZ 85721 USA}
\affiliation{Department of Physics, University of Maryland, College Park, MD 20742 USA}
\author{Jessica N. Howard}
\email[Email address: ]{jnhoward@kitp.ucsb.edu}
 \affiliation{University of California, Santa Barbara, Santa Barbara, CA USA}%
 \affiliation{Kavli Institute for Theoretical Physics, Santa Barbara, CA USA}
\author{Fei Huang}
\email[Email address: ]{fei.huang@weizmann.ac.il }
\affiliation{Department of Particle Physics and Astrophysics, Weizmann Institute of Science, Rehovot 7610001, Israel}
\author{Yuan-Zhen Li}
\email[Email address: ]{yuanzhen.li@uclouvain.be}
\affiliation{Centre for Cosmology, Particle Physics and Phenomenology (CP3),\\ UCLouvain, Louvain-la-Neuve B-1348, Belgium}
\author{Brooks Thomas}
\email[Email address: ]{thomasbd@lafayette.edu}
\affiliation{Department of Physics, Lafayette College, Easton, PA 18042 USA}


\begin{abstract}
One effective way of learning about the production and properties of dark matter 
in the early universe is by extracting information about the primordial dark-matter 
phase-space distribution from the matter power spectrum.  Several years ago a 
simple empirical formula was introduced which successfully reproduces most of the salient features 
of the primordial dark-matter phase-space distribution from the matter power spectrum --- even 
in situations in which this distribution is non-thermal, multi-modal, or exhibits 
other complicated features.  Continuing this line of research, we investigate the extent 
to which machine-learning techniques can improve upon this analytic approach.  
Interestingly, we find that a one-dimensional convolutional neural network not only succeeds 
in reconstructing the dark-matter phase-space distribution with greater accuracy, but 
can also be applied to a broader range of matter power spectra.
\end{abstract}

\maketitle

\tableofcontents


\section{Introduction}


The nature of the dark matter in our universe remains one of the most 
persistent mysteries of contemporary physics.  The scope of theoretical possibilities 
for what the dark matter might be is vast.  Experimental efforts to distinguish between these
possibilities by searching for evidence of non-gravitational interactions between 
the dark matter and the particles of the Standard Model (SM) --- efforts which include searches 
at colliders and beam-dumps, direct-detection experiments, indirect-detection experiments, 
and dedicated axion-detection experiments --- are extensive and ongoing 
(for recent reviews, see, \eg, Refs.~\cite{Hooper:2018kfv,Lin:2019uvt,Slatyer:2021qgc,
Cooley:2022ufh,Cirelli:2024ssz,Bozorgnia:2024pwk,Yu:2025rez}).  To date, these 
efforts have yielded no conclusive signal of non-gravitational interactions between 
the dark and visible sectors.

Despite this fact,
there exist additional astrophysical/cosmological probes which can yield information about 
the dark matter even in situations in which interactions between the dark and visible sectors
are purely gravitational.  One of the most important sources of such information  
is the spatial distribution of matter within our universe.  This distribution is 
typically characterized in Fourier space by the matter power spectrum $P(k)$, 
where $k$ is the wavenumber of the Fourier modes of density perturbations.
Within the context of the ${\Lambda}$CDM cosmology, wherein the dark matter
is perfectly cold, the growth of any particular mode is completely determined 
by the time at which it enters the horizon until the time at which nonlinear effects set in.  
By contrast, in alternative cosmological scenarios, a variety of effects can modify the manner 
in which modes within different ranges of $k$ grow or are suppressed, 
thereby modifying $P(k)$ (for recent reviews, 
see Refs.~\cite{Allahverdi:2020bys,Bechtol:2022koa,Batell:2024dsi}).

Free-streaming is one of the most important such effects.  Dark-matter particles with 
non-negligible velocities can stream out of overdense regions, thereby suppressing the 
formation of structure on distance scales smaller than their particle horizons. 
The impact that this has on $P(k)$ depends on the fraction of the overall population 
of dark matter particles which free-stream at different distance scales --- distance 
scales which correspond roughly to different values of $k$.  This free-streaming 
fraction is essentially determined by the primordial phase-space distribution $f(p)$ of 
the dark-matter particles, which in turn depends on the manner in which these particles 
were initially produced.  As a result, the matter power spectrum 
contains a wealth of information about how the dark matter was initially produced.  Moreover,
since the impact of dark-matter free-streaming on structure formation is purely gravitational
in origin, this is the case even in scenarios wherein dark matter interacts with the SM sector 
only gravitationally.  Given this, it is important to consider how to extract the information 
about $f(p)$ which is encoded within the shape of the matter power spectrum. 

The process of ``inverting" $P(k)$ to constrain $f(p)$ is a challenging 
and technically ill-posed undertaking.  Indeed, the map between $P(k)$ and $f(p)$ is not
one-to-one.  Nevertheless, in a recent paper~\cite{Dienes:2020bmn}, it was shown that salient 
features of $f(p)$ can be extracted from $P(k)$ through a simple heuristic formula.  
This procedure was then extended to the 
non-linear regime --- thereby enabling a similar reconstruction of $f(p)$ from the halo-mass 
function --- in Ref.~\cite{Dienes:2021itb}.  Nevertheless, while the accuracy afforded by
this procedure is sufficient to extract the locations of features in $f(p)$ from $T^2(k)$ 
and the abundances associated with these features, it is not sufficient to uncover the 
detailed profiles of these individual features.

Insight into how the accuracy of these reconstructions might be improved can be gained by
considering approaches that have been developed in order to tackle a similar 
inverse problem which arises in the context of collider searches for new elementary 
particles.  There, the underlying parton interactions under study are morphed and distorted 
by non-perturbative physical processes and detector effects. 
As a result, the final observed data does not perfectly probe the underlying interactions.
Traditional searches forward-simulate parton-level interactions in order to create synthetic data 
under the SM-only and Beyond-the-SM (BSM) hypotheses.  These are then subsequently 
compared to data.  
However, recent work has used machine learning (ML) to instead ``invert" these 
effects via a process commonly called {\it unfolding} (for reviews, see, \eg, 
Refs.~\cite{Cowan:2002in,Blobel:2011fih,Brenner:2019lmf,Canelli:2025ybb}). 

In this paper, we take inspiration from these advances in collider phenomenology and investigate 
the extent to which ML can likewise improve the prospects for unfolding information about $f(p)$ 
from $P(k)$.  This task is more straightforward than its collider analog since the correspondence 
between $f(p)$ and $P(k)$ is more robust --- at least within the linear regime --- than the 
correspondence between the underlying parton-level processes and detector-level events in a 
collider context.  Moreover, there exist software packages, such as 
\texttt{CLASS}~\cite{Lesgourgues:2011re,Blas:2011rf,Lesgourgues:2011rg,Lesgourgues:2011rh},
which can reliably be used to generate pairs of $f(p)$ and $P(k)$ for use in 
supervised training.  

Ultimately, we shall find that a ML-based method for reconstructing $f(p)$ 
from $P(k)$ based on convolutional neural networks (CNNs) significantly
outperforms the analytic method formulated in Ref.~\cite{Dienes:2020bmn}.  
In cases in which $f(p)$ is simple and unimodal, the performance gain is often an order of 
magnitude.  Moreover, even in cases in which $f(p)$ is complicated and/or multimodal, we  
find that the performance gain is quite significant.  There are also certain 
situations~\cite{Dienes:2020bmn, Dienes:2021itb} which the analytic methods were 
not designed to handle.   We nevertheless find that our ML-based approach can 
handle a broad scope of such cases with no apparent difficulty.

We emphasize that the task of reconstructing the {\it primordial}\/ dark-matter velocity 
distribution from $P(k)$ is fundamentally different from, though of course complementary 
to, the task of mapping out the {\it present-day}\/ dark-matter phase-space distribution in 
galaxies and clusters from observational data --- a task to which ML has also recently been 
fruitfully applied~\cite{Putney:2024tfq,Kalda:2025tso,Putney:2025mch,3089976}.
As such, this work represents an application of ML to cosmology as opposed to astrophysics.

The paper is organized as follows.  In Sect.~\ref{sec:theproblem} we give a precise statement 
of the problem we face in attempting to extract $f(p)$ from $P(k)$ and review the progress that 
has been made along these lines using purely analytic methods.  In Sect.~\ref{sec:CNN}, 
we then provide a description of the architecture of the CNN and the data that we use in 
training the network.  In Sect.~\ref{sec:results}, we apply the trained CNN to a variety of 
$P(k)$ functions and demonstrate the network reconstructs $f(p)$ with significantly greater 
accuracy than the heuristic reconstruction formula of Ref.~\cite{Dienes:2020bmn} in all cases.  
In Sect.~\ref{sec:add_features}, we highlight several additional features of our ML-based 
reconstruction procedure.  In Sect.~\ref{sec:conclusions}, we conclude with a summary of our 
results and a discussion of several possible directions for future work.


\section{The goal: Reconstructing $f(p)$ from $P(k)$\label{sec:theproblem}}


In this section, we review the fundamental cosmological problem we wish to address 
in this paper.   We  then summarize previous work in this direction --- work which 
we shall later take as a benchmark when evaluating the efficiency of our new results.

\subsection{Statement of the problem}

In general, any given physical model of dark matter and its production results 
in a prediction for the dark-matter phase-space distribution function $f(p)$.
In writing our full phase-space distribution function $f(\vec x,\vec p,t)$ in the 
simplified form $f(p)$ we are of course assuming spatial isotropy and homogeneity;  
we are also implicitly assuming that this quantity is evaluated at a specific time 
which we may take to be the present time $t_{\rm now}$.  Indeed, for many purposes 
it is often useful to define the corresponding ``comoving phase-space distribution''
\begin{equation}
      g(p)~\equiv~ \frac{a^3p^3}{2\pi^2} f(p)~,
  \label{fg}
\end{equation} 
which is related to the comoving dark-matter number density $\mathcal{N}$  
\begin{equation}
  \mathcal{N} ~\equiv ~\int_{-\infty}^{\infty} d\log p~g(p)~.
  \label{comovingN}
\end{equation}

While $g(p)$ directly connects to an underlying theory, it is the corresponding 
matter power spectrum $P(k)$ which is potentially (although indirectly) observable.  
Indeed, $P(k)$ is often expressed in terms of the so-called {\it squared transfer function},\/ 
$T^2(k) \equiv P(k)/P_{\rm CDM}(k)$, where $P_{\rm CDM}(k)$ is the matter power spectrum 
that would have arisen in a theory assuming purely cold dark matter.   Thus, in terms 
of $T^2(k)$, we see that structure formation is suppressed (or enhanced) if 
$T^2(k)<1$ (or $>1$).

Our interest in this paper therefore boils down to understanding the relationship between 
$g(p)$ and $T^2(k)$.  In principle, for different dark-matter phase-space distributions 
$g(p)$, the linear matter power spectrum $P(k)$ is calculated by solving a full set of 
Boltzmann equations that describe the time-evolution of the perturbed energy density, 
pressure, energy flux, and shear stress in a Friedmann-Robertson-Walker (FRW) cosmology 
in the presence of dark matter having a phase-space distribution function $g(p)$.  This 
calculation can be done using the public code \texttt{CLASS}, which solves the Boltzmann 
equation under various cosmological parameters and energy contents.  Indeed, \texttt{CLASS} 
can perform such calculations while treating the unperturbed present-day phase-space 
distribution $g(p)$ of dark matter as an input.

Thus, the full numerical pipeline constitutes a forward mapping
\begin{equation}
  \mathcal{F}:~~ g(p) \longrightarrow T^2(k)\,.
\end{equation}
There is certainly much that we can learn from studies of this forward map.   For example, 
by starting with a set of different candidate dark-matter models, we may evaluate the 
transfer function $T^2(k)$ for each and compare the results with data.  In this way, we can 
hope to determine which of these models might come closest to making predictions that match 
the observed data.  There has indeed been considerable work in this 
direction~\cite{Bode:2000gq,Viel:2005qj, Boyarsky:2008xj,
Cyr-Racine:2015ihg,Vogelsberger:2015gpr,Konig:2016dzg,Murgia:2017lwo,Irsic:2017ixq,
Murgia:2018now,Kamada:2019kpe,DEramo:2020gpr,Dienes:2021cxp,Huang:2023jxb,
DEramo:2025jsb,DEramo:2025fvy,Zhao:2026wxi}.

More powerful, however, would be an approach which actually {\it extracts}\/ the underlying 
dark-matter phase-space distribution $g(p)$ directly from $T^2(k)$ without the need to make 
and then test a series of guesses.   Indeed, what we really would like to understand is
the {\it inverse}\/ map:
\begin{equation}
  \mathcal{F}^{-1}:~~ T^2(k) \longrightarrow g(p)\, .
\label{eq:inverse_map} 
\end{equation}
Having a handle on this inverse map would allow us to use observational data on $T^2(k)$ 
(such as from, \eg, Ref.~\cite{Planck:2018nkj,Munoz:2019hjh}) in order to extract the 
underlying phase-space distribution for the dark matter and thereby learn about the physics 
of the early universe.

Technically, understanding this inverse map $T^2(k) \rightarrow g(p)$ is an ill-posed 
initial-value problem.  Indeed, it is possible that multiple different phase-space distributions 
may lead to very similar transfer functions.  This situation can be further exacerbated once 
uncertainties in actual astronomical measurements are taken into account.  Nevertheless, the 
need to tackle this inverse problem is fundamental to assessing the implications of cosmological 
data, and is reminiscent of the necessity to interpret the data which emerge from the Large Hadron 
Collider (LHC).~  Indeed, if successful, understanding this inverse map would provide 
a model-agnostic probe of the properties of dark matter in the early universe.

Developing a direct understanding of the inverse map $\mathcal{F}^{-1}$ is notoriously difficult, 
as the forward mapping $g(p)\to T^2(k)$ involves a series of non-linear cosmological integrations.
The traditional approach is to exploit the free-streaming wavenumber
\begin{equation}
  k_{\rm FSH} ~\equiv~ \left[\int_{t_{\rm prod}}^{t_{\rm now}}dt 
    ~\frac{\expt{v(t)}}{a(t)}\right]^{-1}\,,\label{eq:kFSH}
\end{equation}
where $t_{\rm prod}$ is the production time of the dark-matter particles and where
$\expt{v(t)}$ is their average velocity (which redshifts with the scale factor $a(t)$).   
Note that $k_{\rm FSH}$ merely sets a {\it threshold}\/ in $k$-space --- a threshold
below which $T(k)$ is essentially unaffected by the free-streaming of dark-matter particles,
and above which $T(k)$ is dramatically suppressed.
However, while undoubtedly useful for characterizing the effect of free-streaming in cases in which the 
$f(p)$ effectively consists of a single, relatively narrow peak, this approach is 
insensitive to the detailed shape of $g(p)$ --- the very quantity which we would hope to 
uncover.  As a result, this approach can fail --- sometimes dramatically --- in cases in which 
$g(p)$ is broader or multi-modal.

\subsection{An analytical breakthrough: A heuristic reconstruction\label{sec:heuristic}}

A major breakthrough towards solving this inverse problem was achieved in 
Refs.~\cite{Dienes:2020bmn,Dienes:2021itb}.  One of the key insights of this 
method is to express the momentum-space distribution $f(p)$ as a corresponding 
distribution in wavenumber space by promoting the threshold relation in 
Eq.~\eqref{eq:kFSH} into a full-fledged mapping between $p$ and $k$ for each momentum $p$:
\begin{equation}
  k ~=~ \xi\left[\int_{t_{\rm prod}}^{t_{\rm now}}dt 
    ~\frac{p(t)}{a(t)\sqrt{p^2(t)+m^2}}\right]^{-1}\,
  \label{kdef}
\end{equation}
where $m$ is the mass of the dark-matter particle and where $\xi$ is an as-yet-undetermined 
$\mathcal{O}(1)$ numerical factor.  Given this definition for $k$,  we can then convert our 
phase-space distribution $g(p)$ into a corresponding profile $g_k(k)$ in $k$-space by treating 
the substitution of $p$ with $k$ as a standard change of variables:
\begin{equation}
  g_k(k)~=~ \abs{\frac{d p}{d k}} g(p)~.
  \label{Jacobian} 
\end{equation}

We stress that there is no physical justification for treating a particular value of the 
quantity $k$ in Eq.~(\ref{kdef}) as corresponding to a particular momentum $p$.  Indeed, as 
stressed above, for any value of $p$ the corresponding value of $k$ as defined above is 
merely a {\it threshold}\/ value that indicates which wavenumber modes can or cannot be 
affected by physics at momentum scale $p$.  As a result, the entire notion of treating the 
replacement of $p$ with $k$ as an algebraic {\it map between a momentum variable and a 
wavenumber variable}\/ has no physical underpinning.  Following the analysis in 
Refs.~\cite{Dienes:2020bmn, Dienes:2020bmn}, we shall nevertheless make this association 
and see where it leads.   

One immediate consequence of establishing this map is that we now have a new phase-space 
quantity $g_k(k)$ which is a function of a wavenumber $k$ which we may identify as the same 
wavenumber appearing in $T^2(k)$.  This identification is a second critical conceptual leap, 
since it therefore allows us to place $g_k(k)$ and $T^2(k)$ on the same footing as different 
functions living in the same space.   Indeed, these two functions can now even be plotted on 
the same axis.

All of this would be for naught if there were still no way to relate $T^2(k)$ and $g_k(k)$.   
However, given this framework, the authors of Ref.~\cite{Dienes:2020bmn} discovered a heuristic 
relation between $T^2(k)$ and $g_k(k)$.   In particular, this relation states that for a large 
class of transfer functions $T^2(k)$, the underlying phase-space distribution $g_k(k)$ that 
gives rise to it can be fairly accurately extracted through a simple one-line algebraic 
relation~\cite{Dienes:2020bmn} 
\begin{equation}
  g_k(k) ~\approx~  \frac{1}{2} \left( \frac{9}{16} + \left| 
    \frac{d \log T^{2}}{d \log k} \right| \right)^{-1/2} \left| 
    \frac{d^{2} \log T^{2}}{\left( d \log k \right)^{2}} \right|\,.~
\label{eq:reconstruction_analytic}
\end{equation}
Indeed, once $g_k(k)$ is obtained, the original phase-space distribution $g(p)$ is found to 
follow from the Jacobian $|dp/dk|$ in Eq.~(\ref{Jacobian}) with $\xi\approx 5/3$, whereupon 
$f(p)$ can be extracted via Eq.~(\ref{fg}).  Indeed, as discussed in Ref.~\cite{Dienes:2020bmn}, 
it has been found that this simple approximation in Eq.~(\ref{eq:reconstruction_analytic}) 
captures the salient features of the original phase-space distribution, {\it even when the 
dark-matter phase-space distribution exhibits complex shapes such as multiple peaks and 
extended humps}\/.  This heuristic but analytic reconstruction therefore provides an 
unprecedented direct bridge from the observable $T^2(k)$ to the dark-matter phase-space 
distribution $f(p)$ --- one which is relatively straightforward and easy to implement.
Indeed, we shall see explicit examples of results based on this heuristic formula throughout 
this paper.

One important property of the heuristic formula in Eq.~(\ref{eq:reconstruction_analytic})
is that it exhibits {\it $k$-locality}\/.  This refers to the fact that the value of $g_k(k)$ 
at any value of $k$ depends on the behavior of the $T^2(k)$ function only at the same value 
of $k$.  Indeed, the value of $g_k(k)$ at any specific value of $k$ does not depend on the 
behavior of $T^2(k)$ for any {\it other}\/ values of $k$.  We shall return to this point at 
several junctures throughout the rest of this paper.   We stress, however, that at this stage 
$k$-locality is only  a property of the heuristic formula in Eq.~(\ref{eq:reconstruction_analytic}).
In particular, this does not imply that the true reverse map from $T^2(k)$ back to $g_k(k)$ exhibits 
a strict $k$-locality, and indeed one of the goals of this paper will be to examine the extent to 
which our learned CNN solution also exhibits this property.   Of course, the manifest success of 
the heuristic formula suggests that any successful ML-based reconstruction procedure is also likely to 
exhibit $k$-locality to a large extent.

A similar idea that underlies the formulation of the heuristic formula in 
Eq.~(\ref{eq:reconstruction_analytic}) is the expectation that the value of $g_k(k)$ at a 
particular $k$-value $k_\ast$ should not affect the behavior of $T^2(k)$ for any $k<k_\ast$.  
This expectation reflects the underlying idea that dark-matter particles with a given
free-streaming length should not be able to affect the formation of structure on length scales 
exceeding that free-streaming length ({\it i.e.}\/, beyond their particle horizons).  Unlike 
$k$-locality, any sensible theory of dark-matter-induced structure formation must respect the 
limitations imposed by such horizon thresholds exactly.  This observation will also play an 
important role in what follows.

Despite its success in reproducing broad-stroke features, this heuristic result does have 
certain limitations.  For example, this result for $g_k(k)$ is not completely accurate --- there 
can still be certain features within the true underlying dark-matter phase space distribution 
which are not well reproduced using this result.  For this reason alone, it may prove 
advantageous to have a more accurate means of reconstructing $f(p)$ [or equivalently $g_k(k)$].

That said, there are also two other limitations that restrict the broad use of this result.
The first of these limitations stems from the fact that the conceptual framework that led to 
this heuristic result~\cite{Dienes:2020bmn} implicitly assumes that $\log T^2$ is always 
concave down as a function of $k$, or maintains constant slope as a function of $\log k$. 
In other words, the heuristic result implicitly assumes that
\beq 
     \frac{d^2 \log  T^2}{(d \log k)^2}
      ~\leq ~0~~~~ {\rm for~all~} k~.
\label{assumption1}
\eeq 
While this assumption holds in many physical scenarios, 
it is known to break down in some important cases.  For instance, when the true dark-matter 
phase-space distribution $f(p)$ has two widely separated peaks --- for example, one peak 
corresponding to a warm dark-matter (WDM) component and another corresponding to a CDM component 
(whose contribution would therefore be infinitely far away on the $\log\,k$-axis), the transfer 
function $T^2(k)$ reaches a plateau for $k$ values much larger than those associated with the WDM
peak~\cite{Boyarsky:2008xj}.  Furthermore, a slight oscillatory behavior arises in multi-modal 
scenarios at $k$-values slightly larger than the location of a sufficiently narrow
peak~\cite{Dienes:2020bmn,Dienes:2021itb}.  Such behavior inevitably drives the 
transfer function concave-up, and the effect only becomes more prominent as the width of the peak 
decreases.  Of course, a workaround solution might be  to impose a Heaviside theta function 
(or a $\min$ function) that sets $g_k(k)=0$ whenever  the transfer function starts to become 
concave-up.  However, this is an \emph{ad hoc} fix that does not resolve the fundamental difficulty.

The second implied assumption in the heuristic formula takes the form of a limitation on the 
magnitude of the {\it first}\/ derivative of $T^2(k)$, namely
\beq 
    \frac{d \log T^2}{d\log k} ~\geq~ -5/2~~~~ {\rm for~all~} k~.
\label{assumption2}
\eeq
This assumption follows from certain details in Ref.~\cite{Dienes:2020bmn} concerning the 
derivation of Eq.~(\ref{eq:reconstruction_analytic}) --- details which ultimately stem from 
the self-consistency requirement that the so-called ``hot-fraction'' 
function~\cite{Dienes:2020bmn} not exceed unity.  In most cases of physical interest, the 
condition in Eq.~(\ref{assumption2}) is easily satisfied, with potential violations occurring 
only when $T^2(k)$ is sufficiently small that other effects (such as dark-matter acoustic 
oscillations) become dominant.  However, there do exist cases in which these expectations 
can be evaded, whereupon the restriction in Eq.~(\ref{assumption2}) can play an important role.

These restrictions on the universal applicability of the heuristic formula in 
Eq.~(\ref{eq:reconstruction_analytic}) therefore motivate the development of a more robust 
method --- one that can learn the inverse mapping $\mathcal{F}^{-1}$ from concrete examples 
and accurately reconstruct phase-space distributions even in regimes where assumptions inherent 
in the heuristic formula break down.


\section{Reconstruction through machine learning: This is CNN \label{sec:CNN}} 


In this section, we begin the process of tackling this inverse problem through ML-based methods. 
In particular, as discussed in Sect.~\ref{sec:theproblem}, we seek to learn a functional 
which approximately inverts \texttt{CLASS}, \ie, a functional $\mathcal{F}^{-1}_{\rm ML}$ 
which extracts the underlying dark-matter phase-space distribution $g_k(k)$ from the transfer 
function $T^2(k)$ to which it leads. Explicitly, following Eq.~(\ref{eq:inverse_map}), we may write
\begin{equation}
  \mathcal{F}^{-1}_{\rm{ML}}:  ~~~ T^2(k)
    ~\rightarrow~ \gkNN{k} \approx \gktrue{k}~,
\end{equation}
where we henceforth explicitly distinguish between the ``true'' underlying dark-matter 
phase-space distribution $g_k(k)$ and the reconstructed distribution $\hat g_k(k)$ which 
we hope approximates it.

\subsection{Network architecture and physical motivation}  

In Ref.~\cite{Dienes:2020bmn}, it was shown that the heuristic reconstruction formula in
Eq.~(\ref{eq:reconstruction_analytic}) is remarkably successful in extracting the most 
significant features of the dark-matter phase-space distribution $g_k(k)$ from the transfer 
function $T^2(k)$.  Moreover, as discussed in Sect.~\ref{sec:heuristic}, we see from 
Eq.~(\ref{eq:reconstruction_analytic}) that the heuristic formula relies almost 
exclusively on derivative information, which is strictly local in $k$.  Given this 
success, we wish to choose an architecture which emphasizes local features. This is an 
example of an \emph{inductive bias} placed on the ML architecture. Therefore, we shall 
consider a one-dimensional convolutional neural network (CNN) architecture which 
emphasizes $k$-local features of $\logTsq{k}$ while still allowing for connections 
between far-separated $k$-values.  This is crucial as we have seen in 
Sect.~\ref{sec:heuristic} that a number of effects may in principle lead to a 
soft violation of strict $k$-locality. 

\begin{figure*}[ht]
    \centering
    \includegraphics[width=\linewidth]{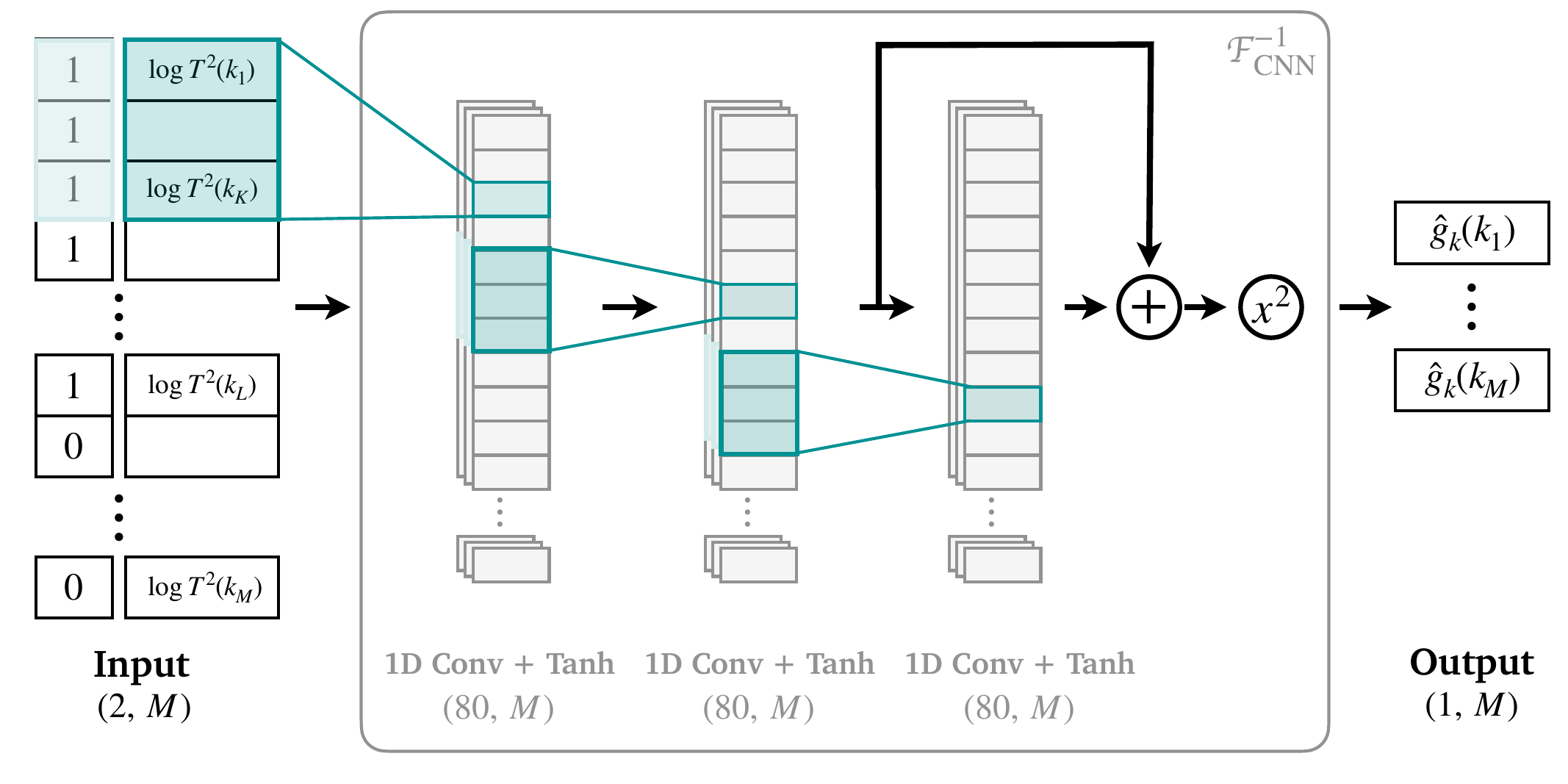}
    \caption{ 
    Sketch of the CNN architecture used in this work.
    Our initial input has two channels. One channel comprises $M = 200$ values of $\logTsq{k}$.  
    For different values of $L$ determined in part by the behavior of 
    the $T^2(k)$ distribution at large $k$, the last $M-L$ values are truncated 
    and forward-filled to simulate observing only part of the $\logTsq{k}$ curve.  
    The other channel is a binary indication of which input values have 
    been truncated. The exact value of $L$ is varied randomly during training. There are three 
    internal convolution layers with $\tanh$ activation functions. The kernel for each has size 
    $(N_{\rm ch}, K)$ where $K=19$ and where $N_{\rm ch}$ is the number of channels in the previous 
    layer (\ie, two for the first hidden layer and 80 for the rest). The output of the third hidden 
    convolution layer is residually combined with the output of the second. This is then squared 
    to produce the final $\hat{g}_k(k)$ predictions.  The loss function $\mathcal{L}$ is evaluated 
    over only non-truncated values --- \ie, using 
    $\{\hat{g}(k_1), \hat{g}(k_1), \ldots, \hat{g}(k_L)\}$.  In order to keep the dimension $M$ 
    constant throughout, the edges are padded with replicated values.} 
    \label{fig:CNN}
\end{figure*}

Our optimized architecture is illustrated schematically in Fig.~\ref{fig:CNN}.  The input
consists of two channels.  The first channel comprises $M$ values $\logTsq{k_i}$
of the squared transfer function evaluated at particular wavenumbers $k_i$, where
$i = 1,2,\ldots,M$.
The second channel contains a binary mask $m_i \in \{0, 1\}$ which is used in the 
implementation of a truncation procedure to be discussed below.  The network itself 
consists of three hidden  one-dimensional convolution layers, each utilizing $80$ channels and a 
broad kernel size of $19$ to capture extended features.  In order to facilitate stable 
gradient flow across 
the network, we incorporate a residual connection bypassing the third convolution layer. 
All hidden layers employ $\tanh(x)$ activation functions. Finally, in order to enforce 
the physical requirement that the phase-space distribution $g_k(k)$ must be strictly 
non-negative, each of the elements $x(k_i)$ of the output array obtained from 
the residual combination of the second and third layers is squared.  It is these squared 
output-array elements which we interpret as the values $\hat g_k(k_i) \equiv x^2(k_i)$ 
of the reconstructed dark-matter velocity distribution through use of our loss function 
(to be discussed below).  We note that it is standard practice to use the squaring 
operation (rather than, \eg, the absolute-value operation) in order to enforce 
non-negativity because the squaring operation preserves smoothness and differentiability.

\subsection{Loss function and optimization}  

Since \texttt{CLASS} allows us to generate pairs $\{\gktrue{k_i}, \logTsq{k_i} \}$
which we shall take to represent truth, with $T^2(k)$ accurately representing the 
transfer function associated with the dark-matter distribution $g_k(k)$, we perform 
supervised regression.  We assess the degree to which
each $\hat g_k(k)$ distribution reconstructed during this process accords with the true
distribution $g_k(k)$ according to a Mean-Squared-Error (MSE) loss function
\begin{equation}
  \mathcal{L}_{\rm MSE} ~\equiv~ \frac{\sum_{i=1}^M m_i 
      \bigl| \gkNN{k_i} - \gktrue{k_i} \bigr|^2}{\sum_{i=1}^M m_i}~.
\label{eq:mse}
\end{equation}
This definition ensures that contributions come from only those $k_i$ values for which $m_i=1$ --- 
\ie, for only those values allowed by the binary mask in the second input channel.

In order to further encourage the learning of physically realistic, smooth $\hat g_k(k)$ 
distributions and to suppress high-frequency numerical noise, we modify $\mathcal{L}_{\rm MSE}$ 
by introducing a secondary penalty term that imposes a penalty for curvature at each 
point.  Since we are dealing with a discretum of $\hat g_k(k_i)$ values, the curvature is
approximated by the second-order finite difference between $\hat g_k(k_i)$ and its nearest 
neighbors.  Our total loss function $\mathcal{L}$, including this penalty term, is thus given by
\begin{equation}
  \mathcal{L} ~=~ \mathcal{L}_{\rm MSE} 
    +\frac{\lambda_{s}}{M} 
    \sum_{i=2}^{M-1} \bigl| \gkNN{k_{i+1}} - 2\gkNN{k_i} + \gkNN{k_{i-1}} \bigr|^2 \, ,
\label{eq:loss}
\end{equation}
where $\lambda_{s}$ is a hyperparameter which the controls the severity of this smoothing penalty.
We train the network for $10^4$ epochs using the \texttt{AdamW}~\cite{Loshchilov:2017bsp} 
optimizer combined with a \texttt{StepLR} learning-rate scheduler which adjusts the learning rate 
$\eta = \eta_0 \gamma^{\lfloor t/\Delta t \rfloor}$ where $t$ is the number of epochs elapsed 
during the training processes, $\gamma$ is the decay rate, $\Delta t$ is the step size (which 
we choose to be $3000$), $\eta_0$ is the initial learning rate, and $\lfloor x\rfloor$ 
denotes the floor function of $x$.  The \texttt{AdamW} optimizer 
implements a weight-decay penalty which is applied directly to the model parameters during 
optimization and whose severity is controlled by a hyperparameter $\lambda_w$.  In order to ensure 
optimal performance without manual bias, we systematically explore the remaining possible 
hyperparameter configurations using the \texttt{Optuna} framework~\cite{akiba2019optuna}.
We find that the performance of the network is optimized for a hyperparameter 
configuration with an initial learning rate $\eta_0 \approx 1.13 \times 10^{-4}$, 
a weight-decay hyperparameter $\lambda_w \approx 2.16 \times 10^{-5}$, a 
smoothing-penalty hyperparameter $\lambda_{s} \approx 9.41 \times 10^{-4}$, and a scheduler 
decay rate $\gamma \approx 0.897$.

\subsection{Training data}  

In order to ensure that our CNN learns a robust and generalizable mapping and does 
not instead merely memorize specific features, we construct a comprehensive training 
dataset spanning the range from simple analytical forms to highly complex and physically 
realistic distributions. For all generated distributions, the true $g_k(k)$ is passed 
through a forward \texttt{CLASS} simulation using standard cosmological parameters in 
order to generate the corresponding transfer function $T^2(k)$.  We then evaluate both 
this $T^2(k)$ function and the corresponding $g_k(k)$ distribution at $M = 200$ 
values of $k$ which are evenly spaced on a logarithmic scale within the 
range $k \in [10^{-2.5}, 10^{3.5}] \, h/{\rm Mpc}$.  Here and throughout this work, 
the wavenumber $k$ is implicitly expressed in units of $h/{\rm Mpc}$ whenever logarithms 
are evaluated, such that $\log k \equiv \log (k / [h/{\rm Mpc}])$.

After sampling $T^2(k)$ and $g_k(k)$ in this manner along such a one-dimension grid of $k$ values,
we implement two measures designed to address two complications which can lead to overfitting.
The first such complication is that $T^2(k)$ typically exhibits severe acoustic oscillations 
once $k$ becomes sufficiently large that the vast majority of dark-matter particles are 
capable of free-streaming.  In order to prevent the CNN from artificially overfitting $\hat g_k(k)$ 
to these highly suppressed, oscillatory features, we introduce a physical cutoff $k_{\rm max}$ 
for each individual $T^2(k)$ function.  Values of $T^2(k)$ for wavenumbers $k < k_{\rm max}$ 
are retained, while values of $T^2(k)$ for $k > k_{\rm max}$ are forward-filled with $T^2(k_{\rm max})$.  
We take this $k_{\rm max}$ to be the largest of the discrete $k$ values along our grid for which 
$T^2(k) > 10^{-4}$ for all $k < k_{\rm max}$.  In situations in which $T^2(k)> 10^{-4}$ for all 
values of $k$ within our one-dimensional grid, $k_{\rm max}$ simply defaults to the maximum value 
of $k$ along the grid.

The second complication is grid-locking --- \ie, the possibility that the network could 
artificially overfit $\hat g_k(k)$ to the specific discrete grid of $k$ values that we have 
adopted.  In order to mitigate this effect and to encourage the CNN to learn the underlying 
continuous physical mapping between $T^2(k)$ and $g_k(k)$, we implement a dynamic grid-jittering 
technique during training.  Specifically, within each training epoch, the entire logarithmic 
evaluation grid is globally shifted by a uniformly sampled, microscopic random offset 
$\delta (\log_{10} k) \in [0, 0.01]$.  We find that this grid-jittering technique 
significantly reduces the effect of grid-locking on the $\hat g_k(k)$ distributions obtained 
from the CNN.~

The dataset on which we train the CNN comprises results for both fully synthetic $g_k(k)$ 
distributions and physically motivated $g_k(k)$ distributions obtained from commonly studied 
dark-matter production mechanisms.  We build the synthetic portion of this dataset using 
mixtures of $N$ log-normal distributions of the form
\begin{equation}
  g_k(k) ~=~ C\,\sum_{n=1}^N\frac{A_n}{\sqrt{2\pi\sigma^2_n}} 
    \,e^{-(\log k -\mu_n)^2/(2\sigma_n^2)}~,
  \label{eq:LogNormal}    
\end{equation}
with logarithmic means $\mu_n$ (which parametrize the average momentum of the 
particles associated with each log-normal peak), logarithmic widths $\sigma_n$ 
(which parametrize the corresponding velocity dispersions), and relative normalization factors 
$A_n$ (which parametrize the corresponding abundances associated with each peak).  The overall
constant $C$ is chosen such that for $\sum_{n=1}^N A_n = 1$ the total present-day abundance of 
dark matter is equal to the value $\Omega_{\rm DM} = 0.264$ inferred from Planck 
data~\cite{Planck:2018vyg}.

Our core dataset includes a large number of cases with differing degrees of complexity:
\begin{itemize}
    \item \textit{Unimodal}\/ ($N=1$): We first consider $g_k(k)$ distributions consisting of a 
    single log-normal peak with $A_1 = 1$.  We include in our training dataset $g_k(k)$ 
    distributions involving all possible combinations that can be made from a set of 21 different 
    $\mu_1$ values within the range $\mu_1 \in [0.23, 4.40]$ and 16 different $\sigma_1$ values 
    within the range $\sigma_1 \in [0.27, 1.13]$.  This set of $\mu_1$ values corresponds
    to a set of average present-day dark-matter velocities $\langle v \rangle$ sampled 
    log-uniformly across the range $\langle v\rangle \in [5 \times 10^{-9}, 5 \times 10^{-7}]$.
    The set of $\sigma_1$ values corresponds to a set of present-day dark-matter velocity 
    dispersions $\sigma_v$ sampled uniformly across the range $\sigma_v \in [0.25, 1.0]$.
    
    \item \textit{Bimodal}\/ ($N=2$): We also consider $g_k(k)$ distributions consisting of 
    two log-normal peaks --- distributions characteristic of scenarios involving two production 
    channels that contribute non-negligibly to the overall dark-matter abundance.
    We include in our dataset $g_k(k)$ distributions with a variety of combination 
    normalizations, peak locations, and widths.  We consider a set of $A_1$ values sampled 
    uniformly within the range $A_1 \in [10^{-4}, 1]$, with $A_2 = 1 - A_1$ in each case.
    We take $\sigma_1 = \sigma_2 = \sigma$ to be equal and consider two different values 
    $\sigma \approx 0.36$ and $\sigma \approx 0.72$ for this common width --- values which 
    correspond to preset-day dark-matter velocity dispersions $\sigma_v = 0.32$ and 
    $\sigma_v = 0.63$, respectively.  We fix the the location $\mu_1$ of the primary peak
    and consider two different values $\mu_1 \approx -0.37$ and $\mu_1 \approx 0.23$ of
    this peak location --- values which correspond to average velocities 
    $\langle v_1\rangle = 10^{-6}$ and $\langle v_1\rangle = 5\times 10^{-7}$ for the 
    portion of the dark-matter abundance associated with this peak.  For each of these $\mu_1$ 
    values, we consider a set of 25 different $\mu_2$ values obtained by sampling the ratio 
    $\langle v_2\rangle /\langle v_1\rangle$ of the average dark-matter velocities associated
    with the two peaks uniformly within the range 
    $\langle v_2\rangle /\langle v_1\rangle \in [10^{-3}, 1]$.
    
    \item \textit{Trimodal}\/
    ($N=3$): We further increase complexity by considering two distinct configurations of 
    three-peak mixtures with $\mu_1 < \mu_2 < \mu_3$.  For both of these configurations, the 
    normalization of the leftmost peak is fixed such that $A_1 = 0.4$, while the abundance of 
    the middle peak is uniformly sampled within the range $A_2 \in [0.1, 0.6]$ and the abundance
    of the third peak is then given by $A_3 = 1 - A_1 - A_2$.  Moreover, for both configurations, 
    the mean $\log k$ value associated with the middle peak is sampled uniformly within the range 
    $\mu_2 \in [-2, 1]$.  In the first configuration, we fix $\sigma_1=\sigma_2=\sigma_3 = 0.25$ 
    and take $\mu_1 = -2.5$ and $\mu_3 = 0.6$.  By contrast, for the second configuration, 
    we fix $\sigma_1 = \sigma_3 = 0.25$ while allowing the width of the central peak to vary. 
    In particular, we sample $\sigma_2$ uniformly within the range $\sigma_2 \in [0.2, 0.5]$.  We
    also take $\mu_1 = -2.2$ and $\mu_3 = 0.4$ for this second configuration such that the
    outer two peaks lie somewhat closer to the middle peak than they do in the first configuration.
    
    \item \textit{Multimodal}\/ ($N \, {\gg}\, 1$): 
    In order to train the network on highly non-trivial distributions, we generate complicated 
    $g_k(k)$ functions by summing a large number of log-normal functions with identical fixed 
    widths $\sigma_n = \sigma$ whose central values $\mu_n$ are uniformly spaced across a specified 
    range $\mu \in [\mu_{\rm min}, \mu_{\rm max}]$  We generate one set consisting of 20 such 
    distributions with $N=80$, $\sigma = 0.25$, $\mu_{\rm min} = -2.5$, and $\mu_{\rm max} = 1$ 
    and another with 8 distributions with $N=30$, $\sigma = 0.2$, $\mu_{\rm min} = -4.1$ and 
    $\mu_{\rm max} = 0.7$.  For both sets of distributions, the normalizations $A_n$ are sampled 
    from a Dirichlet distribution with concentration parameters $\alpha_n = 1$ to ensure a highly 
    randomized, non-uniform mixture.
\end{itemize}

In addition to these synthetic log-normal mixtures, we enrich the training dataset 
with physically realistic $g_k(k)$ distributions with profiles characteristic of the 
freeze-in (FI) and freeze-out (FO) dark-matter production mechanisms.  Rather than relying 
on simple analytical approximations (such as a purely Maxwell-Boltzmann distribution parametrized 
by a single temperature for the FO case), we utilize $g_k(k)$ distributions obtained in 
Ref.~\cite{Du:2021jcj,Huang:2023jxb} which represent exact numerical solutions for the Boltzmann 
equations that govern the evolution of the dark-matter phase-space distribution within both the 
FI and FO regimes.  In particular, we include in our training dataset a set of $g_k(k)$ 
distributions obtained from the infrared freeze-in and from the thermal freeze-out of dark-matter 
particle species with the masses $m \in \{20, 30, 40, 75\}$~keV.

Finally, to each of the $g_k(k)$ distributions described above, we implement one further step in
our data-processing strategy --- one which aligns with the physical expectation that particle 
horizons dictate the distance scales on which particles with non-negligible velocities in 
the early universe can impact the formation of structure.  This expectation,  
which plays a fundamental role in the formulation of the empirical reconstruction 
formula in Eq.~(\ref{eq:reconstruction_analytic}), implies that the manner in which $T^2(k)$
behaves at wavenumbers above any given value of $k$ should not influence $g_k(k)$ at wavenumbers 
below that value of $k$~\cite{Dienes:2020bmn} (see Section~\ref{sec:heuristic}).
However, a number of effects may in principle lead to weak violations of this statement,
especially for values of $k$ near the threshold.  Thus, physically we expect that wavenumbers 
above any given value of $k$ should not strongly influence $g_k(k)$ at wavenumbers below that 
value of $k$. This defines a soft constraint which we wish to place on our model.

In order to incorporate this physical prior into our model, we  apply a random 
truncation procedure during training --- a procedure which is dynamical in the sense that
the location of the truncation varies from one $T^2(k)$ function to the next.  
Explicitly, for each training epoch, we generate $150$ truncated variants of $T^2(k)$ 
for every base sample by selecting a random test cutoff $k_{\rm rand}$ for each variant.
Since this $k_{\rm rand}$ may be either less than or greater than the cutoff $k_{\rm max}$ 
that we impose in order to mitigate the effect of acoustic oscillations, the overall
cutoff $k_{\rm cut}$ for our grid is $k_{\rm cut} = \min\{k_{\rm rand}, k_{\rm max}\}$.  
The input is then structured as a two-channel tensor.  The first channel contains the 
$T^2(k)$ data.  Values of $T^2(k)$ for which $k \leq k_{\rm cut}$ are retained, 
while values for which $k > k_{\rm cut}$ are forward-filled with the $T^2(k)$ value 
that corresponds to the highest value of $k$ below $k_{\rm cut}$.  The second channel acts 
as a binary mask, feeding the network a value $1$ for $k \leq k_{\rm cut}$ and a value $0$ for 
the padded region.

This truncation procedure is profoundly advantageous in a number of respects.  First, 
from a computational perspective, it acts as a powerful data-augmentation technique which 
mitigates overfitting and massively expands the diversity of the training dataset. 
From a theoretical perspective, it encourages the CNN to learn the directional causality 
implied by the existence of particle-horizon thresholds.  From an operational 
perspective, it empowers experimentalists to generate immediate, robust predictions 
for the dark-matter velocity distribution using only partial observational data at 
larger scales without having to wait for future surveys to resolve $T^2(k)$ at 
small scales, where non-linear effects at late times make it exceptionally challenging 
to extract information about the linear matter power spectrum.  That said, we emphasize
that this truncation procedure does not impose or presuppose a directional causality principle 
in the relationship between $T^2(k)$ and $g_k(k)$, but rather softly encourages the CNN to 
learn such a principle.  Thus, were this principle manifestly violated within our 
training dataset, the network would not be artificially thwarted from learning this violation.   

\subsection{Evaluation data}  

In order to rigorously  assess the performance of our CNN, we consider two distinct
classes of $g_k(k)$ distributions within our evaluation dataset.  One of these 
classes comprises distributions which can be used to test the network's capacity to 
\textit{interpolate}, while the other class comprises distributions which can be used 
to test the capacity of the network to \textit{generalize}.

The dataset associated the first of these classes --- the one designed to test
the interpolation capability (\ie, in-domain performance) of the network  ---
comprises $g_k(k)$ distributions qualitatively identical to those in the training dataset, 
but quantitatively distinct.  In particular, we evaluate it on $g_k(k)$ distributions 
which have the same overall mathematical form 
as the distributions used in our training data, but which involve parameter combinations 
that were not used during training.  For example, since our training data includes FI and 
FO distributions with the specific dark-matter masses ($m \in \{20, 30, 40, 75\}$~keV), we
include FI and FO distributions with $m = 10$~keV and $50$~keV in our evaluation dataset.
The former choice of $m$ probes the capability of our ML-based reconstruction procedure for 
mild extrapolation.  By contrast, the latter choice probes its capability for interpolation 
and ensures that the network is genuinely {\it predicting}\/ the form of $\hat g_k(k)$ 
rather than simply memorizing the training examples, although it has learned the general 
morphological characteristics of FI and FO spectra.  In a similar vein, in order to test 
the performance of our CNN on $g_k(k)$ distributions of the form given in 
Eq.~(\ref{eq:LogNormal}) with $N=1$ or $N=2$, we evaluate our reconstruction procedure on 
distributions similar to those described above, but with combinations of $\mu_n$ not used 
in the training data. 

The dataset associated the second of these classes --- the one designed to test
the generalization capability (\ie, out-of-domain extrapolation) of the network  ---
comprises $g_k(k)$ distributions generated using qualitatively different strategies 
than those that were used to generate their training sets.  Since neural networks 
typically extrapolate poorly when applied to such datasets, this represents a highly 
stringent test.  Remarkably, our CNN exhibits impressive generalization capabilities.  
In order to demonstrate this, we evaluate the model against the nine core benchmark 
scenarios considered in Ref.~\cite{Dienes:2020bmn}.  These benchmarks encompass a 
variety of physically motivated single- and multi-component dark sectors with distinct 
cosmological histories, generated using the particular dark-matter production process 
outlined in that work rather than using the Gaussian-mixture approach we have use 
in generating the majority of our training data.


\section{Results}\label{sec:results}


In this section we evaluate the dark-matter phase-space distributions $\hat g_k(k)$ 
extracted through our CNN and compare these results to the true dark-matter phase-space 
distributions $g_k(k)$ which were originally utilized as input to \texttt{CLASS} in 
order to generate the transfer functions $T^2(k)$ on which our CNN operates.  
We also compare these CNN-reconstructed distributions to those which are instead 
reconstructed from the same transfer functions using the heuristic formula in 
Eq.~(\ref{eq:reconstruction_analytic}).  Throughout this section, we shall perform 
these tests in a variety of different cases of increasing complexity.

In order to quantitatively assess the degree to which a given CNN reconstruction is 
more accurate than the corresponding reconstruction resulting from our heuristic 
formula, we define the ``improvement factor''
\begin{equation}
  R~\equiv~\frac{\mathcal{L}_{\rm MSE,h}}{\mathcal{L}_{{\rm MSE}, {\rm CNN}}}~,
  \label{Rdef} 
\end{equation}
where $\mathcal{L}_{\rm MSE,h}$ and $\mathcal{L}_{{\rm MSE}, {\rm CNN}}$ respectively denote
the MSE losses, as defined in Eq.~(\ref{eq:mse}), evaluated for the heuristic and 
ML-based reconstruction procedures, respectively.  Since a smaller MSE loss indicates 
better agreement between $\hat g_k(k)$ and $g_k(k)$, an improvement factor $R > 1$ 
signifies that the trained CNN outperforms the heuristic formula.

\subsection{Unimodal distributions}\label{sec:FI_FO}

\begin{figure}
\includegraphics[width=\linewidth]{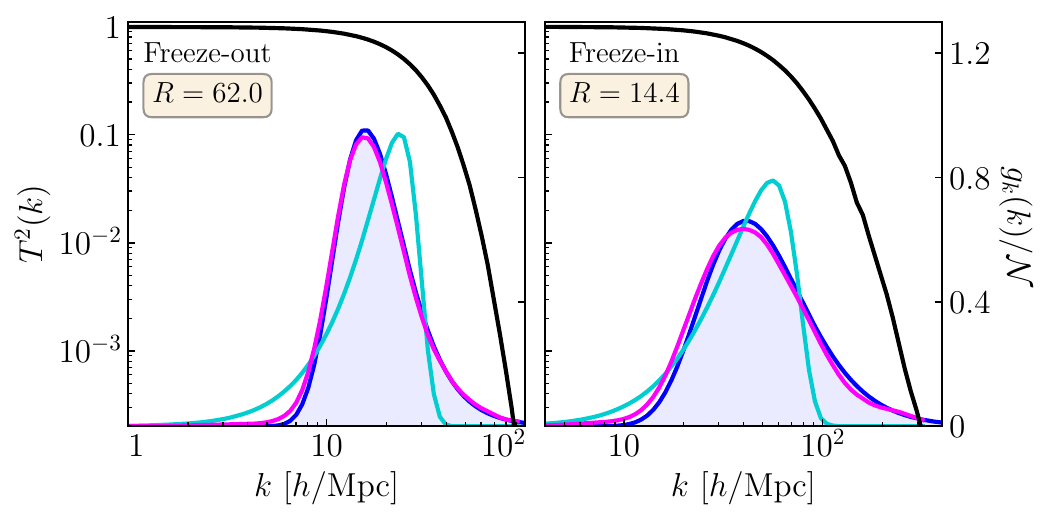} 
    \caption{
    Reconstructing the dark-matter phase-space distribution $g_k(k)$ for unimodal 
    scenarios in which a 50~keV dark-matter candidate is produced via freeze-out 
    (left panel) or freeze-in (right panel).  In each case, the dark blue curve 
    (occassionally hidden behind the magenta curve) represents the true 
    $g_k(k)$ distribution from which the corresponding transfer function (black curve)
    is generated using \texttt{CLASS}.~   The cyan and magenta curves then correspond 
    to the reconstructed distribution functions obtained using the heuristic formula 
    and the CNN, respectively.  The quantity $R$, as defined in Eq.~(\ref{Rdef}), 
    represents the improvement in the MSE that 
    the CNN provides when comparing with the heuristic formula.
\label{fig:unimodal_FIFO}}
\end{figure}

We begin by testing our CNN learned solution on unimodal $g_k(k)$ distributions.  
For the sake of specificity, we apply this learned solution to the FI and FO 
spectra included in our evaluation dataset.
The results for a $m = 50$~keV dark-matter particle are shown in 
Fig.~\ref{fig:unimodal_FIFO}.  In both the FO (left panel) and FI (right panel) cases, 
the heuristic formula successfully predicts a unimodal distribution 
(cyan curve in each panel) whose peak is located at roughly the correct value of $k$.  
However, these reconstructed distributions are tilted 
to the right in both cases, whereas the true distributions (dark blue curve in each panel) 
are each tilted to the left.  In addition, the heuristic formula predicts a relatively
sharp tail on the right side of each peak, while the tails in the true distribution 
always tend more gradually toward zero as $k\to \infty$.
By contrast, the reconstructed distributions $\hat{g}_k(k)$ obtained from the CNN 
(magenta curve in each panel) exhibit a significantly greater accuracy.  Qualitatively,
we observe that the CNN not only correctly reproduces the location and skewness of the peak 
in both the FI and FO cases, but also yields an output $\hat{g}_k(k)$ distribution which aligns 
almost perfectly with the corresponding true $g_k(k)$ distribution.  Quantitatively, 
we observe that the CNN yields improvement factors of $R=14.4$ and $R=62.0$ in the 
FI and FO cases, respectively.  

We have also tested our CNN on a variety of additional unimodal $g_k(k)$ distributions 
beyond the particular examples shown in Fig.~\ref{fig:unimodal_FIFO}, and we find that 
indeed the CNN typically provides an order of magnitude enhancement in the accuracy with 
which the true $g_k(k)$ distribution is reconstructed. 

\subsection{Multimodal distributions}\label{sec:BradyBunch}

\begin{figure*} [ht]
\includegraphics[width=\linewidth]{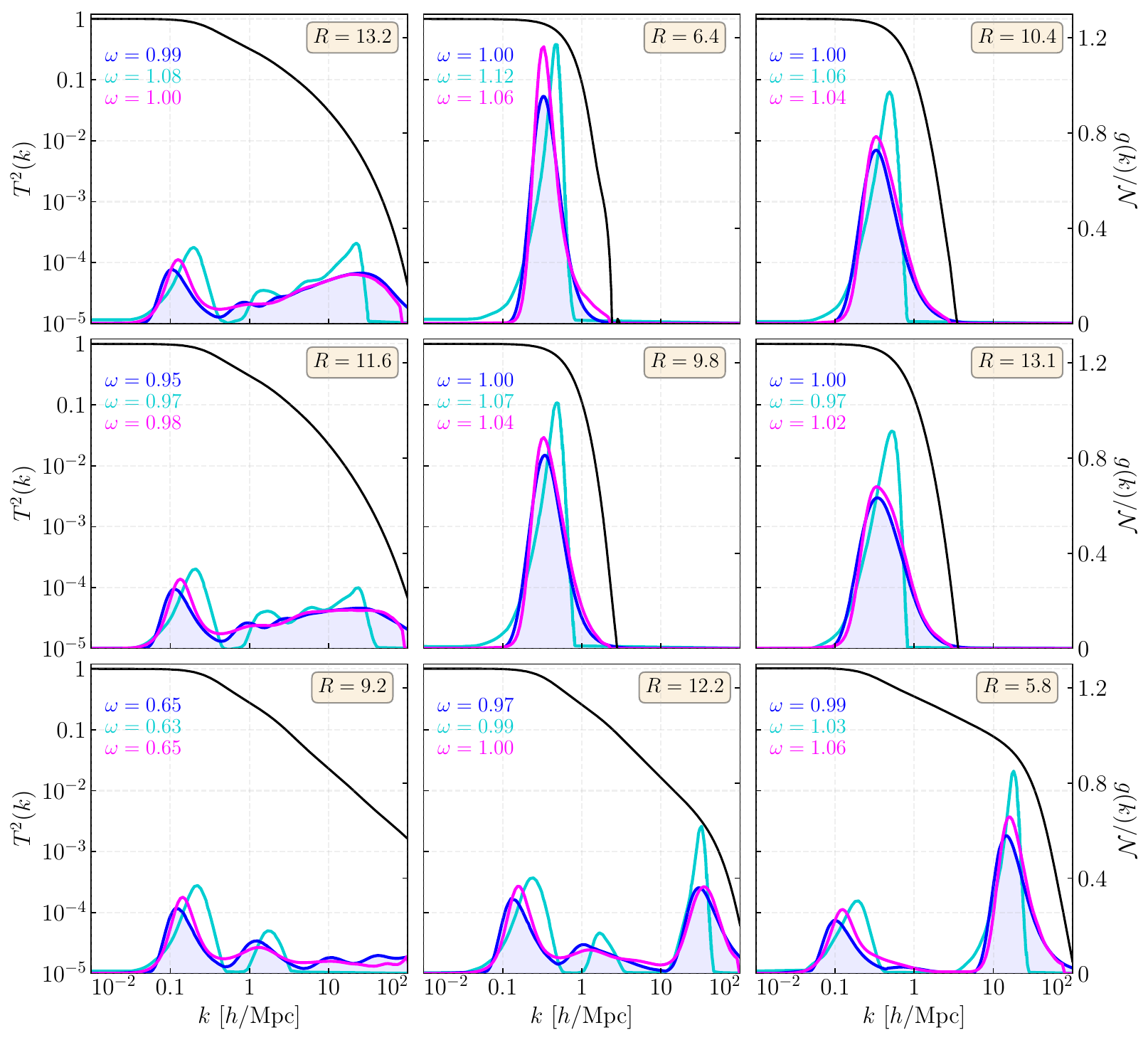}
\caption{Comparisons between the heuristic reconstruction in 
  Eq.~(\ref{eq:reconstruction_analytic}) and our new CNN-based reconstruction, evaluated
  for a variety of complex ``test'' cases drawn from Fig.~18 of Ref.~\cite{Dienes:2020bmn}.    
  In each panel, we show four curves.  The first (blue) is our original dark-matter 
  phase-space distribution which, for the sake of the test, may be regarded as the ``true'' 
  distribution.  We then use this distribution to generate, via \texttt{CLASS}, a 
  corresponding transfer function (black). Given this transfer function, we then calculate 
  two different attempted reconstructions of the original phase-space distribution: 
  one obtained using the analytical heuristic formula (cyan), and another obtained through 
  our CNN (magenta).  We see that in all cases --- including those with unimodal, bi-modal, 
  and even tri-modal distributions, with peaks of assorted heights and widths --- the 
  heuristic formula (cyan) does extremely well in reproducing many of the critical features 
  of the original phase-space distribution (blue)~\cite{Dienes:2020bmn}.  This alone is 
  remarkable, given that the heuristic formula in Eq.~(\ref{eq:reconstruction_analytic}) 
  is a one-line analytical result.  However, we now see that in all of these cases, 
  our CNN can not only do it but do it better, with the CNN significantly outperforming 
  the heuristic approach.   This is especially meaningful, given that these different cases 
  are all not only {\it quantitatively}\/ but even {\it qualitatively}\/ different than 
  those on which the CNN was originally trained.  Indeed, in each case we indicate the 
  ``MSE improvement factor'' $R$ which quantifies how much better than the heuristic 
  formula our CNN performs.  The values of $\omega$ in each case are discussed in 
  Sect.~\ref{sec:locality}.
\label{fig:BradyBunchComparison}}
\end{figure*}

In addition to unimodal distributions, we can also test the trained CNN against more complicated,
multimodal ones.  Given that our training data has already included multimodal distributions 
generated with Gaussian mixtures, we are interested in applying our CNN to cases that in general 
do not belong to a subset of the Gaussian mixtures.  For concreteness, in 
Fig.~\ref{fig:BradyBunchComparison} we exploit the core cases previously considered in 
Ref.~\cite{Dienes:2020bmn}, which cover a broad range of dark-matter phase-space distributions.
Indeed, such distributions are obtained by directly solving a system of Boltzmann equations from 
a model that describes internal decays of a multi-state dark sector.
As we can see from Fig.~\ref{fig:BradyBunchComparison}, the true $g_k(k)$ distributions (blue curves) 
can exhibit many non-trivial features including multiple peaks and trough, extended humps, and large 
gaps between peaks.

When applied to squared transfer functions derived from these $g_k(k)$ distributions, 
we once again find that the CNN reconstructs the original distributions with great 
accuracy, as can be seen from the agreement between the reconstructed $\hat g_k(k)$ 
distributions reconstructed by the CNN (magenta curves) and the true $g_k(k)$ 
distributions.  Despite small differences from the true $g_k(k)$ at various places, 
we find that the CNN even accurately predicts the normalized area 
\begin{equation}
  \omega ~\equiv ~\frac{1}{\mathcal{N}}
    \int_{\log k_-}^{\log k_+} d \log k~g_k(k)\,
  \label{eq:DefOfOmega}
\end{equation}
under the $g_k(k)$ distribution within some window $k_- \leq k\leq k_+$ of interest 
in $k$-space, where $\mathcal{N}$ denotes the co-moving dark-matter number density as 
defined in Eq.~(\ref{comovingN}).  Here and throughout this paper, we take 
$k_- = 10^{-2}~h/{\rm Mpc}$ and $k_+ = 10^2~h/{\rm Mpc}$.  Physically, $\omega$ denotes the 
fraction of the total dark-matter abundance that lies within this range of $k$.  We therefore 
have $\omega = 1$ whenever $g_k(k)$ has support only within of the window $k_- \leq k\leq k_+$ 
and $0\leq\omega <1$ otherwise.

As we see from Fig.~\ref{fig:BradyBunchComparison}, the $\hat g_k(k)$ distribution reconstructed 
by the CNN is accurate not only in cases where $g_k(k)$ has support only within this window, 
but also in cases wherein $g_k(k)$ has support {\it outside}\/ this window as well.
Moreover, we see from the quoted $R$-values that the CNN consistently outperforms the heuristic 
formula, often achieving a result whose MSE loss function is lower by an order of magnitude or 
more.  Moreover, it is also worth emphasizing that while the unimodal $g_k(k)$ distributions 
shown in Fig,~\ref{fig:unimodal_FIFO} are qualitatively similar to $g_k(k)$ distributions included
in our training data, the $g_k(k)$ distributions shown in Fig.~\ref{fig:BradyBunchComparison} 
are qualitatively different from all $g_k(k)$ distributions included in that training data.  
The results shown in Fig.~\ref{fig:BradyBunchComparison} therefore imply that the CNN solution 
is not only capable of interpolating, but also capable of generalizing beyond its training dataset.

\subsection{Stability of learned solutions}\label{app:alpha_beta_gamma}

It is important to test if the learned solution is stable against small perturbations.
We shall first test its stability when applied to unimodal $g_k(k)$ distributions. 
In order to do this, we wish to perturb $T^2(k)$ in a smooth, {\it controlled}\/ way.
Indeed, while one could simply introduce a set of uncorrelated, random perturbations to 
the individual $T^2(k)$ values associated with the $k$ values along our evaluation grid,
the resulting point-to-point fluctuations would modify $d\log T^2(k)/d\log k$ in unphysical
ways and lead to contributions to $\hat g_k(k)$ which are entirely spurious.  In order to 
avoid such spurious contributions to $\hat g_k(k)$, we therefore adopt an alternative approach 
in which we consider a functional form for $T^2(k)$ parametrized by a set of continuous 
parameters. We shall then randomly perturb these parameters around a set of baseline values, 
and examine the effect of these perturbations have on $\hat g_k(k)$.  In particular, we 
consider the parametrization~\cite{Murgia:2017lwo}
\begin{equation}
  T^2(k) ~=~ \big[1+(\alpha k)^\beta\big]^{2\gamma}~,
\end{equation}
where $\alpha,\beta>0$ and $\gamma <0$.  Indeed, in Ref.~\cite{Dienes:2020bmn} it has been 
shown that all transfer functions of this form correspond to strictly unimodal dark-matter 
phase-space distributions.

We take the baseline values for the parameters $\alpha$, $\beta$, and $\gamma$ in our 
stability analysis to be $\{\alpha_0, \beta_0, \gamma_0\} = \{0.01, 2, -4.5\}$.  We then 
generate a family of perturbed $T^2(k)$ functions with different   
$\{\alpha, \beta, \gamma\} = \{\alpha_0 +\Delta\alpha, \beta_0+\Delta\beta, \gamma_0 +\Delta\gamma\}$,
where the perturbations $\Delta\alpha$, $\Delta \beta$, and $\Delta \gamma$ for each 
individual $T^2(k)$ function are randomly sampled uniformly within symmetric intervals 
$-\epsilon |\alpha_0| < \Delta\alpha < \epsilon|\alpha_0|$, 
$-\epsilon|\beta_0| < \Delta\beta < \epsilon|\beta_0|$, and 
$-\epsilon|\gamma_0| <\Delta\gamma < \epsilon |\gamma_0|$ for some $\epsilon$.
For each of these $T^2(k)$ curves, we use the CNN to reconstruct the corresponding 
$\hat g_k(k)$ curve.  We then obtain a variational band around our baseline $T^2(k)$ function 
at each $k$ along our grid by selecting the maximum and minimum values of $T^2(k)$ from among 
our family of perturbed $T^2(k)$ functions  and obtain the corresponding band around our 
baseline $\hat g_k(k)$ distribution in an analogous manner.

\begin{figure*}[ht]
    \centering
\includegraphics[width=0.45\linewidth]{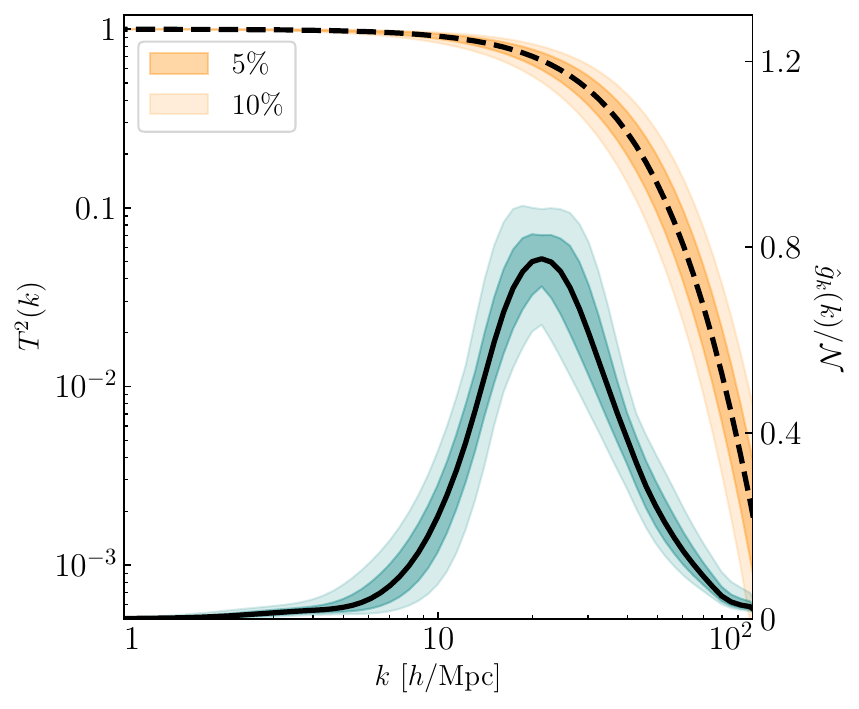}
\includegraphics[width=0.45\linewidth]{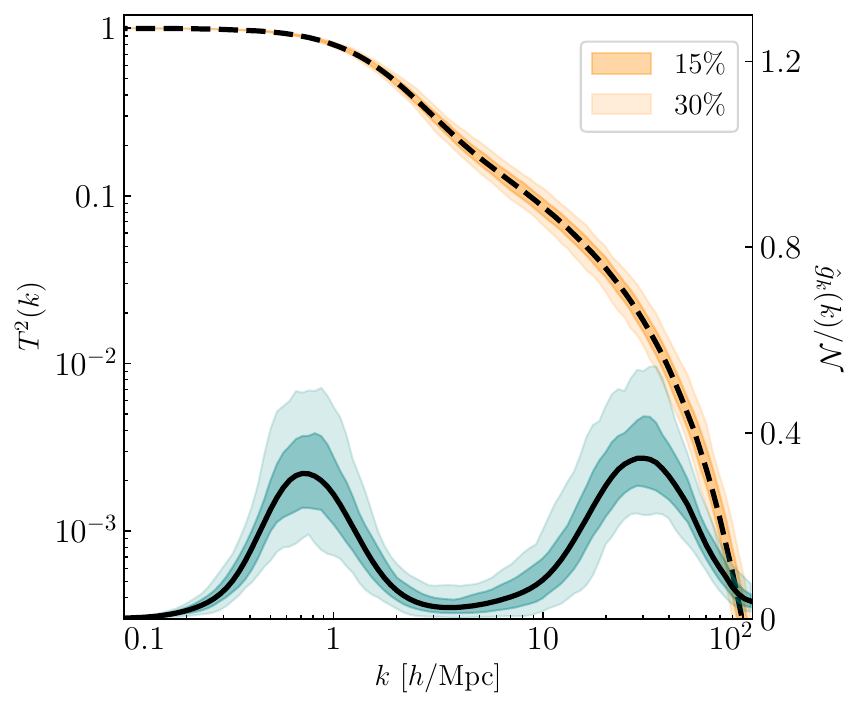}  
  \caption{The effects of varying $T^2(k)$ on the corresponding learned
    CNN solution for $\widehat{g}_k(k)$ for cases in which the true dark-matter 
    phase-space distribution $g_k(k)$ is unimodal (left panel) or bimodal (right panel).  
    Within each panel we show an original CNN input $\logTsq{k}$ (black dashed line) 
    as well as the corresponding CNN output $\hat{g}_k(k)$ (solid black line), while the 
    ``bands'' surrounding these black lines respectively indicate small variations in the 
    input $\logTsq{k}$ (yellow bands) and in the corresponding output $\hat{g}_k(k)$ 
    (teal bands).  We see that small relative changes in  $\logTsq{k}$ lead to small 
    relative changes in $\hat{g}_k(k)$, thereby illustrating that the output of our CNN 
    is qualitatively stable against small variations in its input.
\label{fig:systematicShift_unimodal}}
\end{figure*}

In the left panel of Fig.~\ref{fig:systematicShift_unimodal}, we show the variational bands for 
$T^2(k)$ (yellow) which correspond to $\epsilon = 0.05$ and $\epsilon = 0.1$ --- \ie, to 5\% 
and 10\% variations of $\alpha$, $\beta$, and $\gamma$ around their baseline values --- and the 
corresponding bands for $\hat g_k(k)$ (teal).  The $\epsilon = 0.05$ and $\epsilon = 0.1$ 
bands are constructed using families of 2000 and 4000 different perturbed $T^2(k)$ functions, 
respectively.  We observe that the small variations we have introduced in the $T^2(k)$ 
functions lead to correspondingly small variations in the associated $\hat g_k(k)$ 
distributions reconstructed from these squared transfer functions.  This attests to the  
stability of our ML-based reconstruction procedure against small perturbations when applied
to simple, unimodal $g_k(k)$ distributions. 

In order to assess the stability of this reconstruction procedure when applied to more 
complicated $g_k(k)$ distributions, we perform a similar analysis for bimodal distributions.  
Rather than introduce an explicit parametrization for the corresponding transfer functions, 
we proceed in generating perturbed $T^2(k)$ functions by first evaluating the logarithmic 
difference $\Delta \log T_0^2(k_i) \equiv \log T_0^2(k_{i+1}) - \log T_0^2(k_{i})$ between 
the values of the baseline squared transfer function $T_0^2(k)$ evaluated at each pair of 
adjacent wavenumbers $k_i$ and $k_{i+1}$ along our grid.  We then perturb each of these 
$\Delta \log T_0^2(k_i)$ values by modifying them to 
$\Delta \log T^2(k_i) = (1 + \theta_i)\Delta \log T_0^2(k_i)$, where the $\theta_i$ are 
randomly sampled uniformly within the symmetric interval $-\epsilon < \theta_i < \epsilon$.
A perturbed $T^2(k)$ function is then constructed by starting with the value $T_0^2(k_{\rm min})$
of the baseline transfer function at the lowest $k$ value along our grid and stepping forward
using the perturbed logarithmic differences $\Delta \log T^2(k_i)$.  We generate a family 
of transfer functions in this way and use this family of transfer function in order to construct 
variational bands for $T^2(k)$ and $\hat g_k(k)$ in the same way that we did for the unimodal
case. 

In the right panel of Fig.~\ref{fig:systematicShift_unimodal}, we show variational bands for
$T^2(k)$ (yellow) which correspond to $\epsilon = 0.15$ and $\epsilon = 0.3$ --- \ie, to 15\% 
and 30\% variations around the baseline values for the logarithmic differences 
$\Delta \log T_0^2(k_i)$ --- and the corresponding bands for $\hat g_k(k)$ (teal).
The $\epsilon = 0.15$ and $\epsilon = 0.3$ bands are constructed using families of 2000 and 4000 
different perturbed $T^2(k)$ functions, respectively.  Once again we observe that the 
small variations we have introduced in the $T^2(k)$ functions lead to correspondingly 
small variations in the reconstructed $\hat g_k(k)$ distributions.  This attests that  
our ML-based reconstruction procedure is robustly stable against small perturbations even when 
applied to complicated, multi-modal $g_k(k)$ distributions.

\subsection{Beyond the heuristic formula}\label{sec:plateau}

As noted in Sect.~\ref{sec:heuristic}, the heuristic approach has two intrinsic limitations, 
namely those in Eqs.~(\ref{assumption1}) and (\ref{assumption2}).  While the restriction in 
Eq.~(\ref{assumption2}) is of relatively minor importance and can be overcome reasonably 
effectively through straightforward workarounds, the restriction in Eq.~(\ref{assumption1}) 
is more significant, arising in a variety of different physically-motivated scenarios.   
The existence of this restriction means that the heuristic approach cannot be used in such 
cases, and would lead to nonsensical results if na\"{i}vely applied in situations outside
its regime of validity.   

\begin{figure}\label{fig:beyondHeuristic_plateau}
     \centering
     \includegraphics[width=\linewidth]{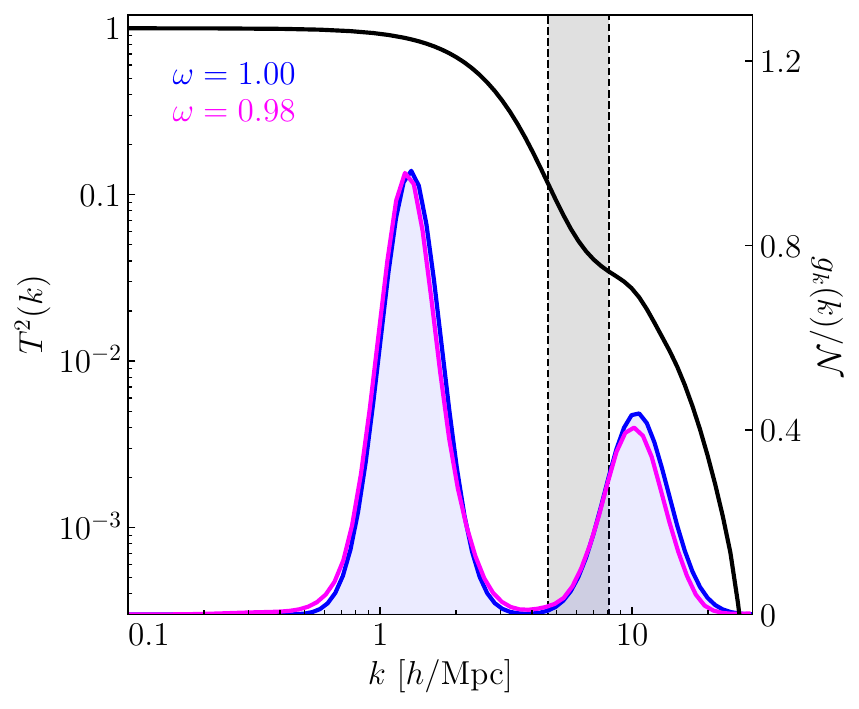}
    \caption{Our CNN reconstruction even performs extremely well in cases where the 
    heuristic formula fears to tread.   Shown in this figure is a case in which the 
    original phase-space distribution (blue) leads to a transfer function (black) which 
    violates one of the critical assumptions --- namely the concave-down condition in
    Eq.~(\ref{assumption1}) --- required for the self-consistency of the heuristic 
    formula.  Indeed, the gray vertical band indicates the region in which this critical 
    assumption is violated.   The presence of such a region is why there is no result 
    to show for the heuristic reconstruction.  However, we see that our CNN has no 
    problem handling cases that include such regions, and leads to a reconstruction 
    (magenta) that continues to reproduce the underlying dark-matter phase-space 
    distribution (blue) with incredible accuracy.
\label{fig:beyond_heuristic}}
\end{figure}

In order to illustrate this, we consider the case shown in Fig.~\ref{fig:beyond_heuristic}, 
where the true $g_k(k)$ distribution is bimodal and exhibits two well-separated, relatively 
narrow log-normal peaks.  In the region between the peaks, dark-matter acoustic oscillations 
give rise to a change in the concavity of $T^2(k)$ within the region of $k$ indicated by the 
the gray-shaded region of the figure.  Since this behavior violates the assumption in 
Eq.~(\ref{assumption1}), we do not expect the heuristic formula to yield reliable results
when applied to $g_k(k)$ distributions of this sort.  Moreover, the assumption in 
Eq.~(\ref{assumption2}) can also be violated for such distributions, since the acoustic 
oscillations can also lead to steep logarithmic slopes $d\log T^2(k)/d\log k$.

By contrast, our CNN can handle such cases without difficulty.  As we can see from 
Fig.~\ref{fig:beyond_heuristic}, the $\hat g_k(k)$ distribution (magenta) 
reconstructed by the CNN closely tracks the profile of the true $g_k(k)$ distribution (blue), 
despite a slight discrepancy between these two distributions at $k$ values above the location
of the maximum for the higher-$k$ peak.  Nevertheless, as indicated by the similarity between
the corresponding $\omega$ values for the two distributions, we observe that our CNN 
predicts the overall abundance associated with the portion of $g_k(k)$ which
falls within the range $k_- \leq k \leq k_+$ impressively accurately.
Thus, we find that our ML-based reconstruction procedure CNN not only outperforms the heuristic 
formula in Eq.~(\ref{eq:reconstruction_analytic}) in scenarios where the latter is applicable, 
but also expands the scope of $T^2(k)$ functions for which the corresponding 
dark-matter phase-space distribution can be reconstructed beyond the regime wherein this
heuristic formula is valid.


\section{Additional features} \label{sec:add_features}


In this section, we discuss several important properties of the heuristic 
reconstruction formula and explore the extent to which these properties are 
also shared by our CNN.~

\subsection{Horizon thresholds\label{sec:thresholds}}

One principle which was fundamental to the formulation
of the heuristic formula in Eq.~(\ref{eq:reconstruction_analytic}) is the expectation
that the value of $g_k(k)$ at a particular $k$-value $k_\ast$ should have no impact 
whatsoever on the shape of $T^2(k)$ for any $k$-value smaller than $k_\ast$.
In other words, colloquially speaking, larger $k$-values for $g_k(k)$ do not affect 
smaller $k$-values for $T^2(k)$.  This expectation reflects the underlying idea that 
dark-matter particles with a given free-streaming length should not be able to affect 
the formation of structure on length scales exceeding that free-streaming length 
(\ie, beyond their particle horizons).

This principle implies that if one had access to $T^2(k)$ for only those $k$-values below 
some threshold value $k_{\rm cut}$, one should nevertheless be able to reconstruct $g_k(k)$ 
at wavenumbers $k < k_{\rm cut}$ just as accurately as one could if one had access to the 
shape of $T^2(k)$ at higher $k$. 

In order to determine whether this threshold behavior is reflected in our CNN learned 
solution, we examine the effect of applying the same truncation procedure 
to our evaluation data that we applied to our training data.  In other words, given 
a $T^2(k)$ function, we not only reconstruct the corresponding 
$\hat{g}_k(k)$ distribution for the full range $k_{\rm min} \leq k \leq k_{\rm max}$ 
of $k$, but we also reconstruct the $\hat{g}_k(k)$ distributions for a number of truncated 
$T^2(k)$ functions obtained from the original distribution by imposing a cutoff
$k_{\rm cut}$ forward-filling the $T^2(k)$ values at $k > k_{\rm cut}$ with 
$T^2(k_{\rm cut})$.  By comparing this $\hat{g}_k(k)$ distributions to the $\hat g_k(k)$
distribution obtained for the full $T^2(k)$ functions, we can determine the extent
to which the threshold behavior associated with particle horizons is reflected in 
our CNN learned solution.

\begin{figure}
    \centering
    \includegraphics[width=\linewidth]{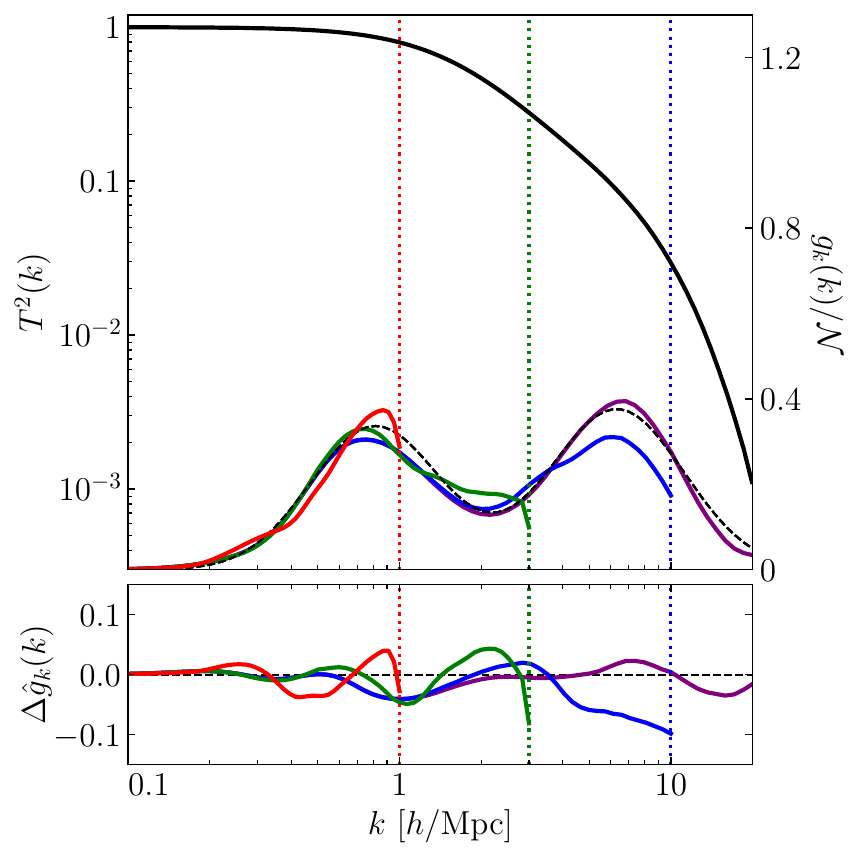}
    \caption{Upper panel: $\hat g_k(k)$ distributions (solid colored curves) reconstructed by the CNN 
    from a $T^2(k)$ function obtained from a bimodal $g_k(k)$ distribution and then 
    truncated above different choices of $k_{\rm cut}$.  The $\hat g_k(k)$ curves shown correspond 
    to the choices $k_{\rm cut} = 1~h/{\rm Mpc}$ (red), $k_{\rm cut} = 3~h/{\rm Mpc}$ (green), 
    $k_{\rm cut} = 10~h/{\rm Mpc}$ (blue), and $k_{\rm cut} = 20~h/{\rm Mpc}$ (purple) --- wavenumbers 
    indicated by the corresponding vertical dashed lines.  The full $T^2(k)$ function 
    (solid black curve) and the $g_k(k)$ distributions from which it was obtained (black dashed curve) 
    are also shown.  Lower panel: the corresponding residuals 
    $\Delta \hat{g}_k(k)\equiv \hat g_{k}(k) - g_k(k)$ for each case.  For $k$ well below $k_{\rm cut}$, 
    the CNN accurately reconstructs the true $g_k(k)$ distribution despite the truncation.  For 
    $k$ near $k_{\rm cut}$, edge effects lead to discrepancies in the reconstruction.}
    \label{fig:horizon_threshold}
\end{figure}

In the upper panel of Fig.~\ref{fig:horizon_threshold}, we show the $\hat{g}_k(k)$ distributions 
(solid colored curves) reconstructed by the CNN from a set of truncated $T^2(k)$ functions 
ultimately obtained from a bimodal $g_k(k)$ distribution.  For the true $g_k(k)$ distribution
(dashed black curve), we have taken $\mu_1 \approx 0.69$ and $\mu_2 \approx 5.31$ (corresponding to 
average velocities $\langle v_1\rangle = 10^{-6}$ and $\langle v_2\rangle = 10^{-7}$ for the
individual log-normal peaks), widths $\sigma_1 = \sigma_2 = 0.72$ (corresponding to a velocity 
dispersion $\sigma_v =  0.63$ for each peak), and normalizations $A_1 = 0.464$ and 
$A_2 = 1-A_1 = 0.536$.  The reconstructed $\hat{g}_k(k)$ distributions correspond to the 
choices $k_{\rm cut} = 1~h/{\rm Mpc}$ (red), $k_{\rm cut} = 3~h/{\rm Mpc}$ (green), 
$k_{\rm cut} = 10~h/{\rm Mpc}$ (blue), and $k_{\rm cut} = 20~h/{\rm Mpc}$ (purple) for the cutoff.  
The vertical dashed lines indicate the corresponding values of $k_{\rm cut}$.  In the lower panel, 
we show the corresponding residuals $\Delta \hat{g}_k(k)\equiv \hat g_{k}(k) - g_k(k)$.  
We also display the corresponding full $T^2(k)$ function (solid black curve) obtained from 
$g_k(k)$.  

In general, we observe from Fig.~\ref{fig:horizon_threshold} that the $\hat g_k(k)$ distributions 
reconstructed from the truncated $T^2(k)$ functions agree well both with the $\hat g_k(k)$ 
distribution reconstructed from the full $T^2(k)$ function and with the true $g_k(k)$ 
distribution at wavenumbers well below $k_{\rm cut}$.  We also observe that for $k$ near 
$k_{\rm cut}$, edge effects associated with the cutoff lead the $\hat g_k(k)$ reconstructed 
from the truncated $T^2(k)$ to deviate more significantly from $g_k(k)$.  In general, we find 
that these deviations are frequently more pronounced in situations in which $k_{\rm cut}$ lies 
in the middle of a feature in $g_k(k)$ rather than in situations in which it lies between features.  
Moreover, when $k_{\rm cut}$ does lie in the middle of a feature in $g_k(k)$, the deviations are 
typically more pronounced when the majority of the abundance associated with the feature lies at   
wavenumbers $k < k_{\rm cut}$ rather than at wavenumbers $k > k_{\rm cut}$.  Edge effects of this
sort are not unexpected, however, and the fidelity with which the CNN learned solution reconstructs
$\hat{g}(k)$ at $k$ well below $k_{\rm cut}$ attests that the underlying physics associated with 
particle horizons is indeed reflected in this learned solution.   

\subsection{$k$-Locality\label{sec:locality}}

In Sect.~\ref{sec:thresholds}, we found that larger $k$-values for $g_k(k)$ do 
not generally affect smaller $k$-values for $T^2(k)$ --- an observation that was 
ultimately encouraged by our truncation procedure.  However, along similar lines, 
one might also wonder about the extent to which \emph{smaller} $k$-values for 
$g_k(k)$ affect {\it larger}\/ $k$-values for $T^2(k)$.  Indeed, if a given $k$-value 
for $g_k(k)$ affects only nearby $k$-values for $T^2(k)$, either larger or smaller, 
we have an approximate $k$-locality.

In the case of the heuristic formula, a given $k$-value for $g_k(k)$ only affects 
$T^2(k)$ at that same value of $k$.  Likewise, the value of $g_k(k)$ at any value of 
$k$ depends only on the value of $T^2(k)$ at that same value of $k$\/, with no 
dependence on $T^2(k)$ at other values of $k$.  Thus, as discussed in 
Sect.~\ref{sec:heuristic}, the heuristic formula exhibits a {\it strict $k$-locality}.
On the other hand, our CNN may not be strictly $k$-local as it can, in principle, 
communicate information across different locations in $k$-space. 

In this subsection, we shall investigate the extent to which our learned CNN solutions 
respect $k$-locality, either strict or approximate.
In order to describe our procedure for investigating this, let us first 
consider how our CNN tests thus far have been constructed.  When we give a specific 
dark-matter phase-space distribution as input into \texttt{CLASS}, we need to 
provide the full $f(p)$ function, or equivalently the full $g_k(k)$ function, 
across all values of $k$.  Given this data, we then use \texttt{CLASS} to calculate 
the corresponding values of $T^2(k)$ over only an appropriately chosen finite range of 
$k$-values.  For example, for all $g_k(k)$ distributions displayed in 
Fig.~\ref{fig:BradyBunchComparison}, we take this range to be 
$10^{-2} \leq k/[h/\mathrm{Mpc}] \leq 10^{2}$.  The results for $T^2(k)$ within this
restricted range of $k$ are then used in the $g_k(k)$ reconstructions based on either 
the heuristic formula or the CNN.~

It is immediately evident upon inspection that the heuristic formula in 
Eq.~(\ref{eq:reconstruction_analytic}) is completely local in $k$-space: 
the value of $g_k(k)$ for any value of $k$ depends on only the function $T^2(k)$ 
and its derivatives at the same value of $k$.  As a result, for the heuristic 
formula, $k$-locality guarantees that our reconstructed distribution function $g_k(k)$ 
will also lie within the same range as $T^2(k)$.
By contrast, our CNN reconstruction may incorporate information which is more non-local 
in $k$, due to our use of a relatively large kernel ($K=19$) as well as the fact that 
our CNN is deep (three layers).  Thus, if indeed the relationship between $g_k(k)$ and $T^2(k)$
is not precisely local, the solution learned by the CNN can capture the non-local aspects of
this relationship, whereas the formula in Eq.~(\ref{eq:reconstruction_analytic}) does not. 

The values of $\omega$ for the original and reconstructed dark-matter phase-space distributions 
provided in each panel of Fig.~\ref{fig:BradyBunchComparison} together provide one useful measure 
for assessing the extent to which $k$-locality is reflected in our ML-based reconstruction procedure.
When generating the $T^2(k)$ function which corresponds to a given $g_k(k)$ distribution, 
we always supply values of $g_k(k)$ to \texttt{CLASS} across the {\it full}\/ range of $k$, but 
generate values of $T^2(k)$ only within the more restricted range 
$10^{-2} \leq k/[h/\mathrm{Mpc}] \leq 10^{2}$.  We restrict the range of $k$ within which we 
evaluate $T^2(k)$ for practical reasons: not only is evaluating $T^2(k)$ is computationally expensive 
for large $k$, but the results obtained for large $k$ can also be numerically unstable.  However, 
this means that we are supplying only partial information about the full $T^2(k)$ distribution    
to either the heuristic formula or to the CNN when reconstructing the dark-matter phase-space 
distribution.  One might therefore worry that $\hat g_k(k)$ might be distorted, even 
{\it within}\/ the restricted range of $k$, by this lack of information --- especially if the 
relationship between $g_k(k)$ and $T^2(k)$ is truly non-local.  Such distortions can be reflected 
not only in the shape of $\hat g_k(k)$ within the restricted range of $k$, but also in its 
overall normalization --- \ie, the dark-matter abundance which lies within that range of $k$.  
Thus, by comparing the $\omega$ values for $g_k(k)$ and the corresponding $\hat g_k(k)$ distribution 
reconstructed by the CNN, we can assess to what extent to which the $\hat g_k(k)$ distribution 
reconstructed by the CNN {\it within}\/ the restricted range of $k$ depends on the properties of 
$T^2(k)$ {\it outside}\/ that restricted range, and thereby obtain information 
regarding the extent to which whether our ML-based reconstruction procedure is $k$-local.   

It is apparent from the $\omega$-values quoted within the different panels of 
Fig.~\ref{fig:BradyBunchComparison} that in all cases the $\omega$ values associated with
the $\hat g_k(k)$ distributions reconstructed by the CNN are relatively close to the $\omega$ 
values associated with the corresponding true $g_k(k)$ distributions.   This provides 
strong evidence that the reconstructed phase-space distribution within our restricted range of $k$
is almost completely insensitive to the properties of $T^2(k)$ outside this range --- and 
therefore that our ML-based reconstruction procedure --- like the heuristic reconstruction 
procedure --- is essentially $k$-local.  We do observe that in certain cases
the difference between the $\omega$ value obtained from our ML-based reconstruction and the
$\omega$ value for the ``true'' $g_k(k)$ distribution is larger than the difference between
between the $\omega$ value obtained from the heuristic formula and the $\omega$ value for the 
corresponding true $g_k(k)$ distribution.  However, this too is not a surprise: while the 
heuristic reconstruction is strictly $k$-local by inspection, the solution learned by the CNN 
need not have this property and indeed may incorporate a limited degree of non-locality.  Indeed, 
this limited degree of non-locality may be precisely what is required in order to produce 
reconstructions that are even more faithful to their input $g_k(k)$ functions than the heuristic 
reconstructions.  In other words, in such cases the slight loss of $k$-locality predicted by 
the CNN for the inverse process $T^2(k)\to g_k(k)$ might actually come closer to representing 
the true nature of the inversion.


\section{Discussion and conclusions\label{sec:conclusions}}


Extracting information about the dark-matter phase-space distribution $f(p)$ 
from information imprinted on the matter power spectrum $P(k)$ --- or, equivalently,
from the transfer function $T^2(k)$ --- is a challenging endeavor, but one which can 
provide critical insight into the nature of the dark matter.  In this paper, we have 
shown that conceptual advances which follow from the functional map between $p$ and 
$k$ originally posited in Ref.~(\cite{Dienes:2020bmn}) --- a functional map defined 
via Eq.~(\ref{kdef}) --- makes it possible to apply machine learning to this 
reconstruction problem in a straightforward way.  
In particular, guided by the principle of $k$-locality and by the physical thresholds 
associated with particle horizons, we have developed and trained a CNN which is 
capable of both qualitatively and quantitatively outperforming the heuristic
reconstruction formula developed in Ref.~\cite{Dienes:2020bmn}.  Indeed, we find that 
this trained CNN yields a significant improvement in accuracy --- frequently
order-of-magnitude level --- over this heuristic formula when applied to the $T^2(k)$ 
functions associated with a variety of dark-matter production mechanisms, including 
freeze-in production, freeze-out production, and production via extended hidden-sector 
decay chains.  

These findings are not only interesting in their own right, but also suggest a number of 
avenues for further investigation into how the process of reconstructing $f(p)$ from $P(k)$ 
can be further enhanced through use of other modern ML methods.  For example, one could use 
more powerful ML architectures, such as transformers, to obtain a closer fit to truth. 
Using a pre-trained, masked self-attention model would hopefully circumvent the comparatively 
limited amount of training data and would strictly enforce the horizon-threshold condition 
via the causal structure.  Alternatively, one could use more interpretable architectures 
like Kolmogorov Arnold Networks (KANs)~\cite{Liu:2024swq} to symbolically regress a new, 
simple analytic formula~\cite{KAN_Recon}.  The straightforward one-dimensional nature of 
this task makes it ideally suited for analysis via this modern method.


\section*{Acknowledgments}


YZL would like to thank Yang Ma for technical assistance with the 
computing cluster at UCLouvain.
The research activities of KRD are supported in part by the U.S.\ Department of Energy 
under Grant DE-FG02-13ER41976 / DE-SC0009913, and also by the U.S.\ National Science 
Foundation through its employee IR/D program. 
JNH was supported by the National Science Foundation under Grant No. NSF PHY-1748958, 
the Gordon and Betty Moore Foundation through Grant No. GBMF7392, and by a UCSB 
Chancellor's postdoctoral fellowship.  The research activities of FH are supported in 
part by ISF Grant 1784/20 and by MINERVA Grant, Project 7141230301.
YZL acknowledges the support as a Postdoctoral Fellow of the FSR fellowship of 
UCLouvain.  Computational resources have been provided by the supercomputing facilities 
of the Universit\'e catholique de Louvain (CISM/UCL) and the Consortium des 
\'Equipements de Calcul Intensif en F\'ed\'eration Wallonie Bruxelles (C\'ECI) funded 
by the Fond de la Recherche Scientifique de Belgique (F.R.S.-FNRS) under convention 
2.5020.11 and by the Walloon Region.  The research activities of BT are supported in 
part by the U.S.\ National Science Foundation under Grant PHY-2310622.  
Parts of this work were performed at the Kavli Institute for Theoretical Physics, 
supported by the National Science Foundation under Grant No. NSF PHY-2309135, and the 
Aspen Center for Theoretical Physics, which is supported by National Science Foundation 
grant PHY-2210452.  The opinions and conclusions expressed herein 
are those of the authors, and do not represent any funding agencies. 

\pagebreak
\bibliography{references}
\end{document}